\documentclass[referee]{raa}           
\usepackage{graphicx,times}
\usepackage{natbib}
\usepackage{amssymb,amsmath}
\usepackage{longtable}
\bibpunct{(}{)}{;}{a}{}{,}

\usepackage[a4paper=true,dvipdfm=true,pagebackref=true]{hyperref}
\hypersetup{pdftitle = The title of my PDF, pdfauthor = My name, pdfsubject= The subject, pdfkeywords = keyword1 keyword2 keyword3} 
\hypersetup{colorlinks = true, linkcolor = green, anchorcolor = red, citecolor = blue, filecolor = red, pagecolor = red, urlcolor = red}

\begin{document}

   \title{A long-term $UBVRI$ photometric study of the pre-main sequence star V350 Cep}

 \volnopage{ {\bf 2012} Vol.\ {\bf X} No. {\bf XX}, 000--000}
   \setcounter{page}{1}

   \author{Sunay Ibryamov\inst{}, Evgeni Semkov\inst{}, Stoyanka Peneva\inst{}}

   \institute{Institute of Astronomy and National Astronomical Observatory,
Bulgarian Academy of Sciences, 72 Tsarigradsko Shose Blvd.,
1784 Sofia, Bulgaria; {\it sibryamov@astro.bas.bg}\\
   {\small Received ; accepted }
}

\abstract{Results from $UBVRI$ optical photometric observations of the pre-main sequence star V350 Cep during the period 2004$-$2014 are presented. 
The star was discovered in 1977 due to its remarkable increase in brightness by more than 5 mag ($R$). In previous studies, V350 Cep was considered a to be a potential FUor or EXor eruptive variable. Our data suggest that during the period of observations the star maintains its maximum brightness with low amplitude photometric variations. Our conclusion is that V350 Cep was probably an intermediate object between FUors and EXors, similar to V1647 Ori.
}
   \authorrunning{Ibryamov et al.}            
   \titlerunning{A long-term $UBVRI$ photometric study of the pre-main sequence star V350 Cep}  
   \maketitle

%
\section{Introduction}           
\label{sect:intro}

Studies of pre-main sequence (PMS) stars are very important for modern astronomy because they give an opportunity to understand the early stages of stellar evolution, as well as to test stellar evolution scenarios. Depending on their initial mass, young stars pass through different periods of stellar activity. The most prominent manifestations of this activity are changes in the star's brightness with various periods and amplitudes.

Photometric and spectroscopic variabilities are the most common characteristics of PMS stars. The most widely-spread type of PMS objects $-$ T Tauri stars are young, low mass stars ($M \leq 2M_\odot$). Their study began after the pioneering work of \cite{Joy+1945}. The main characteristics of T Tauri stars are their emission spectra and their irregular photometric variability. Some T Tauri stars exhibit strong brightness variations over comparatively short time intervals (days, months) with amplitudes of up to several magnitudes. T Tauri stars are separated in two subclasses: classical T Tauri stars (CTTS) surrounded by massive accreting circumstellar disks and weak-line T Tauri stars (WTTS) without evidence for disk accretion (\citealt{Bertout+1989}).

\newpage
According to \cite{Herbst+etal+1994} photometric variability of WTTS is due to the rotation of the stellar surface covered with large cool spots. The periods of variability in WTTS are observed on time scales of days and with amplitudes up to 0.8 mag in the $V$-band. Variability of CTTS is more complicated: the variability is caused by a superposition of cool and hot surface spots producing non-periodic variations with amplitudes up to 2-3 mag. in the $V$-band.

The large amplitude outbursts of PMS stars can be grouped into two main types, named after their respective prototypes: FU Orionis (FUor; \citealt{Ambartsumian+1971}) and EX Lupi (EXor; \citealt{Herbig+1989}). Both types of stars are probably related to low-mass T Tauri stars with massive circumstellar disks, and their outbursts are generally attributed to a sizable increase in accretion rate from the circumstellar disk onto the stellar surface. The outburst of FUor objects last for several decades, and the rise time is shorter than the decline. EXor objects show frequent (every few years or a decade), irregular or relatively brief (a few months to one year) outbursts with an amplitude of several magnitudes ($\Delta$$V$$\approx$3-5). 

The PMS star V350 Cep is located in the field of the reflection nebula NGC 7129, a region with active star formation. The region is immersed in a very active and complex molecular cloud (\citealt{Hartigan+Lada+1985}; \citealt{Miranda+etal+1993}). The distance to NGC 7129 as determined by \cite{Straizys+etal+2014} is 1.15 kpc.

Variability in V350 Cep was discovered by \cite{Gyulbudaghian+Sarkissian+1977} who compared their photographic observations of NGC 7129 with the Palomar Observatory Sky Survey (POSS) plates. V350 Cep was not seen on the POSS O-plate obtained in 1954 (limit $\sim$ 21 mag) and is slightly above the limit of the E-plate. The measured brightness of the star in 1977 was approximately 17.5 mag in $B$-band and 16.5 mag in $V$-band. Follow-up observations by the Sternberg Astronomical Institute (\citealt{Pogosyants+1991}), Sonneberg and Alma-Ata Observatories, which are in plate archives (\citealt{Gyulbudaghian+1980}), suggest that V350 Cep was below the plate limits before 1970, i.e. it was fainter than 17.5 mag in $B$-band. The spectral class of V350 Cep is defined as M2 by \cite{Cohen+Fuller+1985} and as M0 by \citealt{Kun+etal+2009}.

Photometric observations of V350 Cep (\citealt{Gyulbudaghian+Sarkissian+1978}; \citealt{Hakverdian+Gyulbudaghian+1978}; \citealt{Shevchenko+Yakubov+1989}; \citealt{Pogosyants+1991}; \citealt{Semkov+1993}, \citealt{Semkov+1996}, \citealt{Semkov+1997}, \citealt{Semkov+2002}, \citealt{Semkov+2004a}; \citealt{Semkov+etal+1999}) demonstrated changes of brightness, which are typical for CTTS with an amplitude of about 1.5 mag in the $B$-band. All spectral observations of V350 Cep (\citealt{Gyulbudaghian+etal+1978}; \citealt{Magakian+Amirkhanian+1979}; \citealt{Cohen+Fuller+1985}; \citealt{Goodrich+1986}; \citealt{Miranda+etal+1994}; \citealt{Magakian+etal+1999}; \citealt{Semkov+2004b}; \citealt{Kun+etal+2009}) suggest that its spectrum is similar to the CTTS spectra, including being quite variable, having an emission spectrum and a variable P Cygni profile fot the H$\alpha$ line. Collected photometric data indicate that the rise in brightness began some time before 1970, and the light curve of the star resembles that of a classic FUor star V1515 Cyg (see \citealt{Clarke+etal+2005}).

Section 2 gives information about telescopes and cameras used and data reduction. Section 3 describes the derived results and their interpretation.


\newpage
\section{Observations}

The CCD observations of V350 Cep were performed in two observatories with four telescopes: the 2-m Ritchey-Chretien-Coude (RCC), the 50/70-cm Schmidt and the 60-cm Cassegrain telescopes of the Rozhen National Astronomical Observatory (Bulgaria) and the 1.3-m Ritchey-Chretien (RC) telescope of the Skinakas Observatory\footnote{Skinakas Observatory is a collaborative project of the University of Crete, the Foundation for Research and Technology, Greece, and the Max-Planck-Institut f{\"u}r Extraterrestrische Physik, Germany.} of the University of Crete (Greece).

The observations were performed with seven types of CCD cameras: VersArray 1300B (1340 $\times$ 1300 pixels, 20 $\times$ 20 $\mu m/$pixel size) at the 2-m RCC telescope, Photometrics CH360 (1024 $\times$ 1024 pixels, 24 $\times$ 24 $\mu m/$pixel size) and ANDOR DZ436-BV (2048 $\times$ 2048 pixels, 13.5 $\times$ 13.5 $\mu m/$pixel size) at the 1.3-m RC telescope, SBIG ST-8 (1530 $\times$ 1020 pixels, 9 $\times$ 9 $\mu m/$pixel size), SBIG STL-11000M (4008 $\times$ 2672 pixels, 9 $\times$ 9 $\mu m/$pixel size)  and FLI PL16803 (4096 $\times$ 4096 pixels, 9 $\times$ 9 $\mu m/$pixel size) at the 50/70-cm Schmidt telescope, and FLI PL9000 (3056 $\times$ 3056 pixels, 12 $\times$ 12 $\mu m/$pixel size) at the 60-cm Cassegrain telescope. All frames were taken through a standard Johnson-Cousins set of filters. All data were analyzed using the same aperture, which was chosen to have a 6\arcsec radius (while the background annulus was taken from 10\arcsec to 15\arcsec). All frames obtained with the VersArray 1300B, Photometrics CH360 and ANDOR DZ436-BV cameras were bias subtracted and flat field corrected. CCD frames obtained with the SBIG ST-8, SBIG STL-11000M, FLI PL16803 and FLI PL9000 cameras were dark-frame subtracted and flat-field corrected. As a reference, the $UBVRI$ comparison sequence reported in \cite{Semkov+2002} was used. 

The results from our photometric observations of V350 Cep are given in Table~\ref{Tab1}. The columns of the table contain the date and Julian date (JD) of the observation, $UBVRI$ magnitudes of V350 Cep, telescope and CCD camera used. The mean value of the errors in the reported magnitudes are: 0.01$-$0.02 mag for $I$- and the $R$-band data, 0.02$-$0.04 mag for $V$-band data, 0.02$-$0.05 mag for $B$-band data and 0.04$-$0.07 mag for the $U$-band data.

\section{Results and Discussion}

\begin{figure}
   \centering
   \includegraphics[width=12.0cm, angle=0]{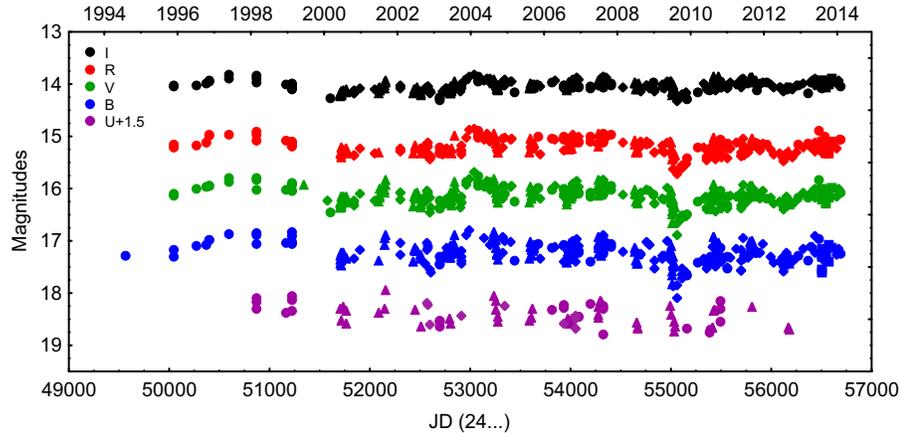}
   \caption{CCD $UBVRI$ light curves of V350 Cep for the period August 1993 $-$ February 2014} 
   \label{Fig1}
   \end{figure}

The presented photometric data are a continuation of our long-term photometric study of V350 Cep. The $UBVRI$ lights curves of V350 Cep from all our CCD observations (\citealt{Semkov+1996}, \citealt{Semkov+1997}, \citealt{Semkov+2002}, \citealt{Semkov+2004a}; \citealt{Semkov+etal+1999} and the present paper) are shown in Figure~\ref{Fig1}. In the figure, circles denote CCD photometric data acquired with the 2-m RCC telescope; triangles $-$ the photometric data taken with the 1.3-m RC telescope; diamonds $-$ the photometric data collected with the 50/70-cm Schmidt telescope, and squares $-$ the photometric data obtained with the 60-cm Cassegrain telescope.

The data reported in the present paper indicate that the brightness of V350 Cep remained close to the maximum value during the period 2004$-$2014 (Table~\ref{Tab1}). Thus, the star has been keeping its maximum brightness during the past 35 yr and for the same period it showed photometric variability with a low amplitude. The observed amplitudes in the period 1993$-$2014 are 0.47 mag for the $I$-band, 0.87 mag for the $R$-band, 1.19 mag for the $V$-band, 1.30 mag for the $B$-band and 0.84 mag for the $U$-band. These values are typical of T Tauri stars surrounded with an accreting circumstellar disk.

\begin{figure}
   \centering
   \includegraphics[width=12.0cm, angle=0]{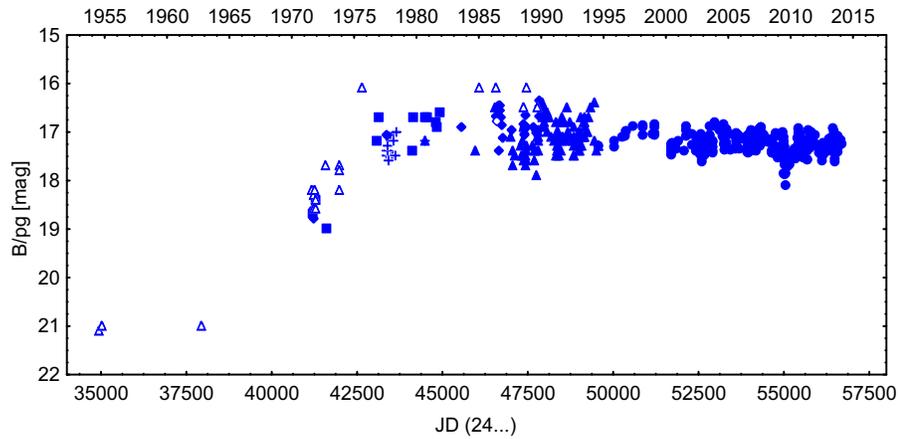}
   \caption{$B/pg$ light curve of V350 Cep from all available observations} 
   \label{Fig2}
   \end{figure}

Figure~\ref{Fig2} shows the long-term $B/pg$-light curve of V350 Cep from all available observations. The circles denote our CCD photometric data (\citealt{Semkov+1996}, \citealt{Semkov+1997}, \citealt{Semkov+2002}, \citealt{Semkov+2004a}; \citealt{Semkov+etal+1999} and the present paper); triangles $-$ the photographic data from the Rozhen Schmidt telescope (\citealt{Semkov+1993}; \citealt{Semkov+1996}); diamonds $-$ the photographic data from \cite{Pogosyants+1991}; squares $-$ the photographic data from the Asiago Schmidt telescope (\citealt{Semkov+etal+1999}); empty diamonds symbols $-$ the photographic data from \cite{Shevchenko+Yakubov+1989}; pluses $-$ the photographic data from Byurakan Schmidt telescope (\citealt{Gyulbudaghian+Sarkissian+1977}; \citealt{Semkov+1993}); the empty triangles $-$ the limit of the photographic data from the POSS plates, the Sternberg Astronomical Institute plate archive \cite{Pogosyants+1991} and the Asiago Schmidt telescope plate archive (\citealt{Semkov+etal+1999}). The available photometric data suggest that the period of strong increase in brightness continued to about 1978 was followed by a period of irregular variability around the level of maximum brightness lasting up to now. 

\begin{figure}
   \centering
   \includegraphics[width=10.0cm, angle=0]{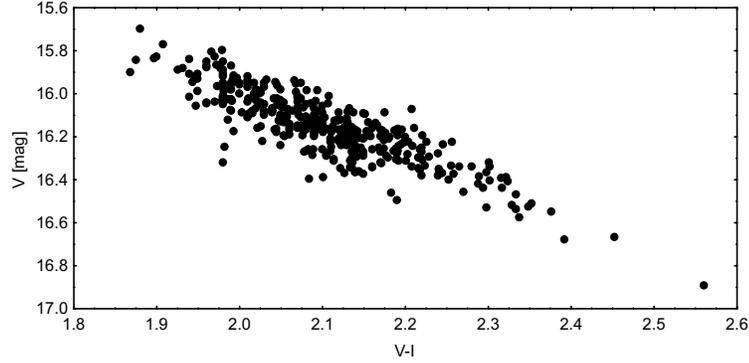}
   \caption{Relationship between $V$ magnitude and $V-I$ color index in the period of all our CCD observations} 
   \label{Fig3}
   \end{figure}

\newpage
Another important result from our photometric study is the variation of color indices with stellar brightness. The measured color index $V-I$ versus the stellar magnitude $V$ during the period of our CCD observations is plotted in Figure~\ref{Fig3}. A clear dependence can be seen from the figures: the star becomes redder as it fades. The other indices $V-R$ and $B-V$ show a similar trend on the color-magnitude diagram. Such color variations are typical of T Tauri stars with large cool spots, whose variability is produced by rotation of the spotted surface. Consequently, V350 Cep shows photometric characteristics of WTTS (variability with small amplitude in a time scale of days). On the other hand, the observed spectra of V350 Cep can be classified as a CTTS spectrum (\citealt{Magakian+etal+1999}; \citealt{Semkov+2004b}). As can be seen from the Table~\ref{Tab1}, V350 Cep shows a very strong ultraviolet excess $-$ a characteristic also typical of CTTS. Moreover, the long-term light curve of V350 Cep is similar to FUor type objects such as V1515 Cyg. These discrepancies make V350 Cep a unique object, which is very difficult to classify.

Regardless of its similarity to T Tauri stars, the large amplitude outburst of V350 Cep can only be explained only as an episode of enhanced accretion. We suggest that V350 Cep is an object similar to V1647 Ori (see \cite{Aspin+Reipurth+2009} and references therein). The eruptive PMS star V1647 Ori was discovered in 2004 during its large amplitude outburst and is considered to be a unique object which shows the photometric characteristics of FUors and spectral characteristics of EXors. Both stars V350 Cep and V1647 Ori show similar photometric and spectral features. These include: a large amplitude outburst continuing several years, random fluctuations in brightness with amplitudes of a few tenths of a magnitude and timescales of several days (\citealt{GarciaAlvarez+etal+2011}), reddening of the color indices with decreasing brightness, connection to reflection nebulae (\citealt{Miranda+etal+1994}), an emission of line spectrum during the maximum light and a variable P Cyg profile for the H$\alpha$ line.   

FUors and EXors have been classified in terms of their wide range of available photometric and spectral properties, but their outbursts are thought to have been caused by an enhanced accretion rate. According to \cite{Aspin+2011}, the viewing inclination angle of the star/disk system can play a significant role in the observed spectral features, and therefore for the classification of the object as FUor or EXor. It is possible that the two types of eruptive variables FUors and EXors could be much closer in nature.
\newpage 
\normalem
\begin{acknowledgements}
This study was partly supported by ESF and Bulgarian Ministry of Education and Science under the contract BG051PO001-3.3.06-0047. The authors thank the Director of Skinakas Observatory Prof. I. Papamastorakis and Prof. I. Papadakis for award telescope time. The research has made use of the NASA Astrophysics Data System Abstract Service.
\end{acknowledgements}
 
\bibliographystyle{raa}
\bibliography{bibtex}

\newpage
\begin{longtable}{lllllllll}
\caption{Photometric CCD observations of V350 Cep during the period August 2004 - February 2014}\\
\hline\hline
\noalign{\smallskip}  
Date \hspace{1.5cm} &	J.D. (24...) \hspace{2mm}	&	I	\hspace{4mm} & R \hspace{4mm} & V \hspace{4mm} & B \hspace{4mm} & U \hspace{4mm} & Telescope \hspace{1mm} & CCD	\\
\noalign{\smallskip}  
\hline
\endfirsthead
\caption{continued.}\\
\hline\hline
\noalign{\smallskip}  
Date \hspace{1.5cm} &	J.D. (24...) \hspace{2mm}	&	I	\hspace{4mm} & R \hspace{4mm} & V \hspace{4mm} & B \hspace{4mm} & U \hspace{4mm} & Telescope \hspace{1mm} & CCD	\\
\noalign{\smallskip}  
\hline
\noalign{\smallskip}  
\endhead
\hline
\label{Tab1}
\endfoot
\noalign{\smallskip}
2004 Aug 11 & 53228.575 &	13.98 &	15.06 &	15.93 &	16.98 & - &	1.3m & Phot\\
2004 Aug 18 & 53235.552 &	13.96 & - &	15.91 &	16.94 & - &	1.3m & Phot\\
2004 Aug 20 & 53238.409 &	13.93 &	14.96 &	15.83 &	16.85 &	16.57 &	1.3m & Phot\\
2004 Aug 22 & 53239.545 &	13.97 & - &	15.95 &	16.97 & - &	1.3m & Phot\\
2004 Aug 23 & 53240.550 &	13.97 & - &	15.91 &	16.96 & - &	1.3m & Phot\\
2004 Aug 23 & 53241.470 &	14.00 & - & 16.00 &	17.04 & - &	1.3m & Phot\\
2004 Sep 08 & 53257.400 &	13.99 &	15.07 &	15.97 &	17.00 &	16.67 &	1.3m & Phot\\
2004 Sep 09 & 53258.329 &	14.04 &	15.14 &	16.08 &	17.14 &	16.84 &	1.3m & Phot\\
2004 Sep 18 &	53266.552 &	14.11 & - &	16.20 &	17.25 & - &	1.3m & Phot\\
2004 Sep 20 &	53268.559 &	14.09 & - &	16.17 &	17.28 & - &	1.3m & Phot\\
2004 Sep 20 &	53269.425 &	14.06 &	15.21 &	16.14 &	17.26 & - &	1.3m & Phot\\
2004 Sep 23 &	53272.470 &	14.03 & - &	16.04 &	17.15 &	- & 1.3m & Phot\\
2004 Sep 28 &	53277.305 &	14.09 &	15.25 &	16.18 &	17.22 &	16.97 &	1.3m & Phot\\
2004 Sep 29 &	53278.316 &	14.12 &	15.31 &	16.27 &	17.28 &	17.06 &	1.3m & Phot\\
2004 Sep 30 &	53279.353 &	14.13 &	15.32 &	16.27 &	17.37 & - &	1.3m & Phot\\
2004 Oct 03 &	53281.561 &	14.10 & - &	16.19 &	17.25 & - &	1.3m & Phot\\
2004 Nov 17 &	53327.377 &	14.03 &	15.22 &	16.20 &	17.23 & - &	Sch &	ST-8\\
2004 Nov 18 &	53328.321 &	13.96 &	15.05 &	15.97 &	16.94 & - &	Sch &	ST-8\\
2004 Nov 20 &	53330.347 &	13.97 &	15.05 &	15.96 &	16.98 & - &	Sch &	ST-8\\
2004 Dec 08 &	53348.289 &	14.03 & - &	16.13 & - & - &	Sch &	ST-8\\
2004 Dec 09 &	53349.271 &	14.01 & - &	16.08 & - & - &	Sch &	ST-8\\
2004 Dec 10 &	53350.243 &	13.92 &	15.03 &	15.95 &	17.01 &	16.76 &	Sch &	ST-8\\
2005 Feb 10 &	53412.219 &	13.96 &	15.07 &	15.98 &	17.04 & - &	Sch &	ST-8\\
2005 Feb 11 &	53413.212 &	13.96 &	15.05 &	15.96 & - & - &	Sch &	ST-8\\
2005 Mar 13 &	53442.596 &	14.16 & - &	16.26 &	17.38 & - &	2m & VA\\
2005 Aug 13 &	53596.417 &	14.10 &	15.31 &	16.27 &	17.35 &	17.03 &	1.3m & Phot\\
2005 Aug 22 &	53605.276 &	13.99 &	15.05 &	15.98 &	17.03 & - &	1.3m & Phot\\
2005 Aug 23 &	53606.279 &	14.03 &	15.15 &	16.07 &	17.12 & - &	1.3m & Phot\\
2005 Aug 26 &	53609.331 &	14.06 &	15.18 &	16.15 &	17.19 & - &	1.3m & Phot\\
2005 Aug 27 &	53610.284 &	14.07 &	15.22 &	16.21 &	17.28 &	16.98 &	1.3m & Phot\\
2005 Aug 29 &	53611.526 &	14.09 & - &	16.21 &	17.29 & - &	1.3m & Phot\\
2005 Sep 14 &	53628.322 &	14.02 &	15.14 &	16.05 &	17.12 &	16.82 &	1.3m & Phot\\
2005 Sep 18 &	53632.461 &	14.07 &	15.19 &	16.15 &	17.20 & - &	1.3m & Phot\\
2005 Sep 19 &	53633.440 &	14.08 &	15.24 &	16.15 &	17.18 & - & 1.3m & Phot\\
2005 Sep 24 &	53638.423 &	14.02 &	15.14 &	16.05 &	17.09 & - &	1.3m & Phot\\
2005 Oct 03 &	53646.521 &	14.04 & - &	16.08 &	17.15 & - &	1.3m & Phot\\
2005 Oct 31 &	53675.393 &	14.01 & - &	16.03 & - & - &	Sch &	ST-8\\
2005 Nov 03 &	53678.201 &	14.04 &	15.02 &	15.99 &	17.07 & - &	2m & VA\\
2005 Nov 25 &	53700.199 &	14.06 &	15.35 &	16.21 &	17.36 & - &	Sch &	ST-8\\
2006 Mar 28 &	53822.594 &	13.97 &	15.05 &	16.04 &	17.14 &	16.83 &	2m & VA\\
2006 Apr 24 &	53849.597 &	13.94 &	15.07 &	15.95 &	17.12 & - &	Sch &	ST-8\\
2006 Jul 19 &	53936.402 &	14.03 &	15.28 &	16.17 &	17.41 &	17.09 &	Sch &	ST-8\\
2006 Jul 22 &	53938.555 &	14.08 & - &	16.13 & - & - &	2m & Phot\\
2006 Jul 23 &	53939.520 &	14.07 &	15.07 &	16.04 &	17.09 &	16.73 &	2m & Phot\\
2006 Jul 24 &	53940.517 &	14.04 &	15.02 &	16.02 &	17.11 &	16.79 &	2m & Phot\\
2006 Aug 21 &	53968.510 &	14.12 & - &	16.31 &	17.43 & - &	1.3m & Phot\\
2006 Aug 22 &	53969.509 &	14.08 & - &	16.25 &	17.36 & - &	1.3m & Phot\\
2006 Aug 29 &	53976.527 &	14.05 & - &	16.19 &	17.27 & - &	1.3m & Phot\\
2006 Aug 29 &	53977.490 &	14.06 & - &	16.19 &	17.29 & - &	1.3m & Phot\\
2006 Aug 30 &	53978.483 &	14.03 &	15.21 &	16.12 &	17.21 &	16.99 &	1.3m & Phot\\
2006 Sep 02 &	53981.343 &	14.01 & - &	16.05 &	17.12 & - & 1.3m & Phot\\
2006 Sep 06 &	53985.209 &	14.02 &	15.18 &	16.07 &	17.17 & &	1.3m & Phot\\
2006 Sep 07 &	53986.201 &	13.99 & - &	16.07 &	17.16 & - &	1.3m & Phot\\
2006 Sep 08 &	53987.443 &	13.99 &	15.13 &	16.04 &	17.12 & - &	1.3m & Phot\\
2006 Sep 09 &	53988.470 &	14.03 & - &	16.12 &	17.23 & - &	1.3m & Phot\\
2006 Sep 11 &	53990.213 &	14.04 & - &	16.11 &	17.19 & - &	1.3m & Phot\\
2006 Sep 12 &	53991.470 &	14.02 & - &	16.07 &	17.19 & - &	1.3m & Phot\\
2006 Sep 14 &	53992.522 &	14.08 & - &	16.21 &	17.31 & - &	1.3m & Phot\\
2006 Sep 14 &	53993.484 &	14.04 & - &	16.15 &	17.26 & - &	1.3m & Phot\\
2006 Sep 17 &	53995.527 &	14.06 & - &	16.18 &	17.34 & - &	1.3m & Phot\\
2006 Sep 18 &	53996.518 &	13.98 & - &	16.00 &	17.05 & - &	1.3m & Phot\\
2006 Sep 22 &	54001.321 &	14.07 &	15.26 &	16.19 &	17.34 &	17.10 &	1.3m & Phot\\
2006 Sep 23 &	54002.406 &	14.06 & - &	16.21 &	17.35 & - &	1.3m & Phot\\
2006 Sep 25 &	54004.388 &	14.03 & - &	16.10 &	17.21 & - &	1.3m & Phot\\
2006 Sep 26 &	54005.386 &	13.98 & - &	16.00 &	17.10 & - &	1.3m & Phot\\
2006 Sep 27 &	54006.312 &	14.02 &	15.18 &	16.08 &	17.17 & - &	1.3m & Phot\\
2006 Sep 29 &	54008.425 &	14.06 & - &	16.17 &	17.27 & - &	1.3m & Phot\\
2006 Sep 30 &	54009.186 &	14.05 &	15.22 &	16.13 &	17.25 &	17.07 &	1.3m & Phot\\
2006 Oct 05 &	54014.237 &	14.06 &	15.24 &	16.16 &	17.22 &	17.00 &	1.3m & Phot\\
2006 Oct 18 &	54027.473 &	13.97 &	15.06 &	16.09 & - & - &	Sch &	ST-8\\
2006 Oct 19 &	54028.416 &	13.94 &	14.98 &	15.92 &	17.04 & - &	Sch &	ST-8\\
2006 Oct 20 &	54029.392 &	13.93 &	15.01 &	15.91 &	17.02 & - &	Sch &	ST-8\\
2006 Nov 17 &	54057.239 &	13.99 &	15.08 &	16.09 &	17.23 &	16.94 &	Sch &	ST-8\\
2006 Nov 18 &	54058.254 &	14.00 &	15.11 &	16.12 &	17.40 &	17.18 &	Sch &	ST-8\\
2006 Nov 19 &	54059.259 &	14.01 &	15.13 &	16.10 &	17.21 & - &	Sch &	ST-8\\
2006 Nov 20 &	54060.202 &	13.98 &	15.07 &	16.07 &	17.27 &	17.01 &	Sch &	ST-8\\
2006 Nov 21 &	54061.429 &	13.88 & - &	15.87 &	16.97 & - &	Sch &	ST-8\\
2006 Dec 14 &	54084.193 &	14.07 &	15.11 &	16.10 &	17.23 &	16.96 &	2m & VA\\
2006 Dec 15 &	54085.244 &	14.06 &	15.13 &	16.09 &	17.20 &	16.96 &	2m & VA\\
2006 Dec 16 &	54086.247 &	13.93 &	15.02 &	15.98 &	17.10 & - &	Sch &	ST-8\\
2007 Apr 09 &	54199.518 &	14.09 &	15.02 &	16.08 & - & - &	2m & VA\\
2007 Apr 09 &	54200.477 &	14.05 &	15.10 &	16.17 &	17.40 &	16.71 &	2m & VA\\
2007 Jun 26 &	54278.359 & - &	14.97 &	15.85 &	16.94 &	16.78 &	1.3m & Phot\\
2007 Jun 30 &	54282.347 &	13.96 &	15.07 &	15.98 &	17.09 & - &	1.3m & Phot\\
2007 Jul 01 &	54283.462 &	13.98 &	15.12 &	16.02 &	17.14 &	16.93 &	1.3m & Phot\\
2007 Jul 02 &	54284.310 &	13.94 & - &	15.94 &	17.05 & - &	1.3m & Phot\\
2007 Jul 03 &	54285.424 &	14.02 &	15.20 &	16.15 &	17.25 &	16.99 &	1.3m & Phot\\
2007 Jul 18 &	54300.446 &	13.93 & - &	15.93 &	17.06 & - &	1.3m & ANDOR\\
2007 Jul 20 &	54301.546 &	13.93 & - &	15.95 &	17.07 & - &	1.3m & ANDOR\\
2007 Jul 23 &	54305.439 &	13.90 &	14.98 &	15.87 &	17.02 &	16.66 &	1.3m & ANDOR\\
2007 Jul 24 &	54306.458 &	13.93 &	15.05 &	15.92 &	17.03 &	16.68 &	1.3m & ANDOR\\
2007 Aug 02 &	54314.557 &	14.03 & - &	16.12 &	17.30 & - &	1.3m & ANDOR\\
2007 Aug 14 &	54327.283 &	14.08 &	15.12 &	16.14 &	17.25 &	17.30 &	2m & VA\\
2007 Aug 15 &	54328.290 &	14.03 &	15.06 &	16.05 &	17.21 & - &	2m & VA\\
2007 Aug 16 &	54329.314 &	14.07 &	15.11 &	16.08 &	17.17 &	16.76 &	2m & VA\\
2007 Aug 17 &	54330.423 &	13.99 &	14.97 &	15.98 &	17.09 &	16.82 &	2m & VA\\
2007 Aug 18 &	54331.340 &	13.89 &	15.01 &	15.85 &	16.90 & - &	Sch &	ST-8\\
2007 Aug 19 &	54332.330 &	13.87 &	14.98 &	15.85 &	16.91 & - &	Sch &	ST-8\\
2007 Aug 20 &	54333.340 &	13.94 &	15.07 &	16.01 &	17.07 & - &	Sch & ST-8\\
2007 Sep 10 &	54354.495 &	13.91 &	15.06 &	15.92 &	17.02 & - &	Sch & ST-8\\
2007 Sep 14 &	54357.511 &	13.89 &	15.03 & - & - & - &	Sch &	ST-8\\
2007 Nov 06 &	54411.260 &	14.04 & - &	16.03 &	17.12 & - &	2m & VA\\
2007 Nov 08 &	54413.291 &	13.99 &	14.98 &	15.97 &	17.06 & - &	2m & VA\\
2008 Mar 01 &	54526.672 &	14.07 &	15.13 &	16.12 &	17.22 & - &	Sch &	STL-11000M\\
2008 Jun 28 &	54646.358 &	13.98 &	15.10 &	16.03 &	17.14 & - &	1.3-m & ANDOR\\
2008 Jun 29 &	54647.339 &	14.01 &	15.16 &	16.11 &	17.22 & - &	1.3-m & ANDOR\\
2008 Jul 06 &	54654.348 &	14.11 &	15.31 &	16.25 &	17.40 & - &	1.3m & ANDOR\\
2008 Jul 08 &	54656.426 &	14.00 & - &	16.13 &	17.18 & - &	1.3m & ANDOR\\
2008 Jul 13 &	54661.383 &	14.08 &	15.27 &	16.27 &	17.44 &	17.08 &	1.3m & ANDOR\\
2008 Jul 24 &	54672.367 &	14.12 &	15.32 &	16.30 &	17.44 &	17.18 &	1.3m & ANDOR\\
2008 Jul 25 &	54673.377 &	14.08 &	15.28 &	16.28 &	17.43 &	17.17 &	1.3m & ANDOR\\
2008 Aug 02 &	54680.544 &	14.00 & - &	16.12 &	17.26 & - &	1.3m & ANDOR\\
2008 Aug 27 &	54706.334 &	14.08 &	15.10 &	16.09 &	17.26 & - &	Sch &	STL-11000M\\
2008 Aug 28 &	54707.374 &	14.14 &	15.21 &	16.20 &	17.29 & - &	Sch &	STL-11000M\\
2008 Oct 21 &	54761.236 &	14.04 &	15.12 &	16.03 &	17.02 & - &	Sch &	STL-11000M\\
2008 Oct 23 &	54763.229 &	14.04 &	15.16 &	16.02 & - & - &	Sch &	STL-11000M\\
2008 Nov 20 &	54791.173 &	14.08 &	15.14 &	16.04 &	17.09 & - &	Sch &	STL-11000M\\
2008 Dec 26 &	54827.342 &	13.98 & - &	16.14 & - & - &	2m & VA\\
2009 Jan 10 &	54842.212 &	14.12 &	15.31 &	16.30 &	17.54 & - &	Sch &	STL-11000M\\
2009 Jan 12 &	54844.256 &	14.15 &	15.32 &	16.29 &	17.38 & - &	Sch &	STL-11000M\\
2009 Mar 24 &	54915.407 &	14.02 &	15.18 &	16.20 &	17.30 & - &	Sch &	STL-11000M\\
2009 Mar 26 &	54917.497 &	14.07 &	15.28 &	16.19 &	17.35 & - &	Sch &	STL-11000M\\
2009 Apr 16 &	54938.435 &	14.10 &	15.31 &	16.23 &	17.29 & - &	Sch &	STL-11000M\\
2009 May 19 &	54971.420 &	13.97 &	15.13 &	16.00 &	17.01 & - &	Sch &	STL-11000M\\
2009 Jun 12 &	54994.504 &	13.95 &	15.11 &	16.09 &	17.21 &	16.76 &	1.3m & ANDOR\\
2009 Jun 18 &	55000.545 &	13.99 & - &	16.12 &	17.19 & - &	1.3m & ANDOR\\
2009 Jun 21 &	55003.502 &	14.04 &	15.23 &	16.22 &	17.28 &	16.93 &	1.3m & ANDOR\\
2009 Jun 24 &	55006.500 &	13.96 &	15.12 &	16.05 &	17.10 & - &	1.3m & ANDOR\\
2009 Jun 29 &	55011.527 &	14.08 &	15.34 &	16.34 &	17.39 & - &	Sch &	FLI\\
2009 Jul 02 &	55014.531 &	14.22 & - &	16.67 &	17.85 & - &	1.3m & ANDOR\\
2009 Jul 07 &	55019.530 &	14.15 &	15.44 &	16.44 &	17.60 & - &	1.3m & ANDOR\\
2009 Jul 10 &	55022.535 &	14.16 & - &	16.51 &	17.67 & - &	1.3m & ANDOR\\
2009 Jul 14 &	55027.440 &	14.13 &	15.48 &	16.42 &	17.51 & - &	Sch &	FLI\\
2009 Jul 15 &	55028.462 &	14.18 &	15.64 &	16.53 &	17.88 & - &	Sch &	FLI\\
2009 Jul 16 &	55029.469 &	14.14 &	15.41 &	16.35 &	17.56 & - &	Sch &	FLI\\
2009 Jul 19 &	55031.517 &	14.02 & - &	16.23 &	17.32 & - &	1.3m & ANDOR\\
2009 Jul 25 &	55037.503 &	14.11 &	15.34 &	16.35 &	17.44 &	17.24 &	1.3m & ANDOR\\
2009 Jul 27 &	55040.440 &	14.14 & - &	16.47 & - & - &	1.3m & ANDOR\\
2009 Jul 28 &	55041.273 &	14.12 &	15.37 &	16.36 &	17.46 &	17.05 &	1.3m & ANDOR\\
2009 Jul 31 &	55044.390 &	14.13 &	15.37 &	16.37 &	17.51 &	17.16 &	1.3m & ANDOR\\
2009 Aug 21 &	55065.354 &	14.33 &	15.73 &	16.89 &	18.10 & - &	Sch &	FLI\\
2009 Aug 22 &	55066.284 &	14.29 &	15.65 &	16.68 &	17.86 & - &	Sch &	FLI\\
2009 Oct 07 &	55112.379 &	14.24 &	15.62 &	16.58 &	17.55 & - &	Sch &	FLI\\
2009 Oct 08 &	55113.354 &	14.21 &	15.55 &	16.54 &	17.65 & - &	Sch &	FLI\\
2009 Oct 09 &	55114.246 &	14.23 &	15.58 &	16.53 &	17.73 & - &	Sch &	FLI\\
2009 Nov 20 &	55156.238 &	14.19 &	15.52 &	16.52 &	17.61 & - &	Sch &	FLI\\
2009 Nov 21 &	55157.267 &	14.17 &	15.55 &	16.55 &	17.62 & - &	Sch &	FLI\\
2009 Nov 25 &	55161.239 &	14.31 &	15.44 &	16.50 &	17.67 &	17.18 &	2m & VA\\
2010 Mar 12 &	55268.553 &	14.18 &	15.22 &	16.26 &	17.42 & - &	2m & VA\\
2010 May 13 &	55330.374 &	14.00 &	15.23 &	16.19 &	17.35 & - &	Sch &	FLI\\
2010 Jun 08 &	55356.418 &	14.03 &	15.27 &	16.16 &	17.36 & - &	Sch &	FLI\\
2010 Jun 10 &	55358.468 &	14.07 &	15.39 &	16.37 &	17.55 & - &	Sch &	FLI\\
2010 Jun 12 &	55359.508 &	14.12 &	15.46 &	16.37 &	17.46 & - &	Sch &	FLI\\
2010 Jun 12 &	55360.442 &	14.09 &	15.43 &	16.41 &	17.40 & - &	Sch &	FLI\\
2010 Jul 13 &	55391.395 &	14.13 &	15.26 &	16.34 &	17.38 &	17.26 &	2m & VA\\
2010 Jul 18 &	55396.339 &	14.10 &	15.29 &	16.40 &	17.32 &	17.22 &	2m & VA\\
2010 Aug 04 &	55413.309 &	14.04 &	15.36 &	16.26 &	17.42 & - & Sch &	FLI\\
2010 Aug 06 &	55415.397 &	14.03 &	15.31 &	16.23 &	17.34 & - &	Sch &	FLI\\
2010 Aug 07 &	55416.360 &	14.03 &	15.18 &	16.11 &	17.13	& - &	Sch &	FLI\\
2010 Aug 11 &	55420.318 &	14.03 &	15.30 &	16.24 &	17.37 &	17.16 &	1.3m & ANDOR\\
2010 Aug 12 &	55421.418 &	14.00 &	15.25 &	16.17 &	17.31 & - &	1.3m & ANDOR\\
2010 Aug 24 &	55433.490 &	13.88 &	15.03 &	15.94 &	16.93 & - &	1.3m & ANDOR\\
2010 Aug 25 &	55434.336 &	13.90 &	15.08 &	15.94 &	16.94 &	16.84 &	1.3m & ANDOR\\
2010 Aug 26 &	55435.370 &	13.92 &	15.10 &	15.99 &	17.03 &	16.84 &	1.3m & ANDOR\\
2010 Sep 08 &	55447.520 &	14.09 &	15.41 &	16.22 &	17.48 & - &	Sch &	FLI\\
2010 Sep 08 &	55448.435 &	14.05 &	15.30 &	16.25 &	17.26 & -	&	Sch &	FLI\\
2010 Sep 09 &	55449.480 &	14.12 &	15.45 &	16.44 &	17.50 & - &	Sch &	FLI\\
2010 Oct 11 &	55481.452 &	13.86 &	15.14 &	16.07 &	17.04 & - &	1.3m & ANDOR\\
2010 Oct 29 &	55499.326 &	13.96 &	15.06 &	16.06 &	17.15 &	16.81 &	2m & VA\\
2010 Oct 30 &	55500.301 &	14.07 &	15.17 &	16.18 &	17.31 &	16.51 &	2m & VA\\
2010 Nov 01 &	55502.276 &	14.06 &	15.19 &	16.21 &	17.34 &	17.07 &	2m & VA\\
2010 Nov 04 &	55505.298 &	14.06 &	15.36 &	16.34 &	17.39 & - &	Sch &	FLI\\
2010 Nov 05 &	55506.313 &	14.08 &	15.41 &	16.39 &	17.41 & - &	Sch &	FLI\\
2011 Jan 01 &	55563.256 &	14.10 &	15.43 &	16.39 &	17.52 & - &	Sch &	FLI\\
2011 Jan 06 &	55568.234 &	14.04 &	15.15 &	16.17 &	17.31 & - &	2m & VA\\
2011 Jan 09 &	55571.275 &	14.08 &	15.25 &	16.34 &	17.54 & - &	2m & VA\\
2011 Feb 06 &	55599.228 &	13.98 &	15.25 &	16.15 &	17.33 & - &	Sch &	FLI\\
2011 Apr 04 &	55656.422 &	14.02 &	15.13 &	16.16 &	17.16 & - &	Sch &	FLI\\
2011 May 22 &	55704.392 &	14.00 &	15.32 &	16.23 &	17.39 & - &	Sch &	FLI\\
2011 May 23 &	55705.335 &	14.07 &	15.39 &	16.39 &	17.58 & - &	Sch &	FLI\\
2011 May 24 &	55706.324 &	14.04 &	15.37 &	16.34 &	17.57 & - &	Sch &	FLI\\
2011 May 25 &	55707.335 &	13.91 &	15.16 &	16.09 &	17.19 & - &	Sch &	FLI\\
2011 Jun 09 &	55722.354 &	13.90 &	15.05 &	15.95 &	16.96 & - &	Sch &	FLI\\
2011 Jun 21 &	55734.470 &	14.02 &	15.38 &	16.32 &	17.35 & - &	Sch &	FLI\\
2011 Jun 24 &	55737.404 &	13.99 &	15.21 &	16.17 &	17.29 & - &	Sch &	FLI\\
2011 Jul 27 &	55770.412 &	13.99 &	15.31 &	16.24 &	17.34 & - &	Sch &	FLI\\
2011 Aug 18 &	55791.563 &	13.88 & - &	15.95 &	17.03 & - &	1.3m & ANDOR\\
2011 Aug 23 &	55797.366 &	13.95 &	15.22 &	16.16 &	17.22 & - &	Sch &	FLI\\
2011 Aug 24 &	55798.390 &	13.96 &	15.21 &	16.09 &	17.14 & - &	Sch &	FLI\\
2011 Aug 25 &	55799.392 &	13.94 &	15.21 &	16.10 &	17.14 & - &	Sch &	FLI\\
2011 Sep 10 &	55815.290 &	13.89 &	15.06 &	15.99 &	17.03 &	16.77 &	1.3m & ANDOR\\
2011 Sep 11 &	55816.444 &	13.89 &	15.05 &	15.95 &	17.01 & - &	1.3m & ANDOR\\
2011 Sep 19 &	55824.310 &	13.97 &	15.18 &	16.17 &	17.24 & - &	1.3m & ANDOR\\
2011 Sep 23 &	55828.302 &	13.94 &	15.17 &	16.03 &	17.16 & - &	Sch	& FLI\\
2011 Oct 13 &	55848.318 &	13.94 &	15.14 &	16.09 &	17.21 & - &	1.3m & ANDOR\\
2011 Oct 30 &	55865.280 &	14.14 &	15.15 &	16.12 &	17.24 & - &	2m & VA\\
2011 Nov 29 &	55895.296 &	14.01 &	15.26 &	16.19 &	17.24 & - &	Sch &	FLI\\
2011 Nov 30 &	55896.277 &	14.00 &	15.27 &	16.20 &	17.23 & - &	Sch &	FLI\\
2011 Dec 29 &	55925.222 &	13.97 &	15.14 &	16.05 &	17.06 & - &	Sch &	FLI\\
2012 Jan 30 &	55957.221 &	13.97 &	15.17 &	16.20 & - & - &	2m & VA\\
2012 Mar 16 &	56003.488 &	14.03 &	15.25 &	16.18 &	17.32 & - &	Sch &	FLI\\
2012 Apr 12 &	56030.484 &	14.08 &	15.36 &	16.30 &	17.43 & - &	Sch &	FLI\\
2012 Jun 12 &	56091.443 &	14.07 &	15.34 &	16.26 &	17.40 & - &	Sch &	FLI\\
2012 Jun 18 &	56096.522 &	14.07 &	15.36 &	16.29 &	17.41 & - &	Sch &	FLI\\
2012 Jul 11 &	56120.447 &	14.11 &	15.41 &	16.29 &	17.34 & - &	Sch &	FLI\\
2012 Jul 13 &	56122.443 &	14.13 &	15.44 &	16.35 &	17.51 & - &	Sch &	FLI\\
2012 Jul 14 &	56123.438 &	14.15 &	15.48 &	16.40 &	17.60 & - &	Sch &	FLI\\
2012 Aug 01 &	56141.439 &	14.11 &	15.37 &	16.34 &	17.47 & - &	1.3m & ANDOR\\
2012 Aug 19 &	56159.393 &	14.07 &	15.37 &	16.20 &	17.31 & - &	Sch &	FLI\\
2012 Aug 20 &	56160.398 &	14.10 &	15.39 &	16.26 &	17.40 & - &	Sch &	FLI\\
2012 Aug 21 &	56161.453 &	14.07 &	15.34 &	16.17 &	17.23 & - &	Sch &	FLI\\
2012 Aug 22 &	56162.378 &	14.08 &	15.35 &	16.26 &	17.33 & - &	Sch &	FLI\\
2012 Sep 02 &	56173.374 &	14.07 &	15.29 &	16.25 &	17.35 &	17.16 &	1.3m & ANDOR\\
2012 Sep 11 &	56182.290 &	14.09 &	15.34 &	16.31 &	17.44 &	17.21 &	1.3m & ANDOR\\
2012 Sep 23 &	56194.364 &	14.08 &	15.37 &	16.27 &	17.36 & - &	Sch &	FLI\\
2012 Oct 09 &	56210.265 &	14.11 &	15.41 &	16.31 &	17.40 & - &	Sch &	FLI\\
2012 Nov 18 &	56250.355 &	14.04 &	15.32 &	16.24 &	17.33 & - &	Sch &	FLI\\
2012 Dec 14 &	56276.305 &	14.03 &	15.16 &	16.18 &	17.30 & - &	2m & VA\\
2013 Feb 04 &	56328.241 &	13.99 &	15.23 &	16.09 &	17.16 & - &	Sch &	FLI\\
2013 Mar 17 &	56369.516 &	14.18 &	15.14 &	16.18 &	17.38 & - &	2m & VA\\
2013 Apr 11 &	56394.389 &	13.98 &	15.20 &	16.14 &	17.13 & - &	Sch &	FLI\\
2013 May 02 &	56415.490 &	13.94 &	15.18 &	16.07 &	17.12 & - &	Sch &	FLI\\
2013 May 30 &	56443.497 &	13.90 &	15.12 &	16.01 &	16.93 & - &	Sch &	FLI\\
2013 May 31 &	56444.481 &	14.03 &	15.29 &	16.18 &	17.24 & - &	Sch &	FLI\\
2013 Jul 04 & 56478.438 & 13.97 & 14.90 & 15.85 & 16.98 & - & 2m & VA\\
2013 Aug 02 & 56507.444 & 14.09 & 15.13 & 16.18 & 17.32 & - & 2m & VA\\
2013 Aug 04 & 56509.368 & 14.04 & 15.32 & 16.20 & 17.33 & - & Sch & FLI\\
2013 Aug 05 & 56510.452 & 13.95 & 15.11 & 16.02 & 17.16 & - & 60cm & FLI\\
2013 Aug 06 & 56511.493 & 13.97 & 15.19 & 16.23 & 17.61 & - & 60cm & FLI\\
2013 Aug 07 & 56512.481 & 14.01 & 15.24 & 16.13 & 17.55 & - & 60cm & FLI\\
2013 Aug 08 & 56513.469 & 13.94 & 15.16 & 16.05 & 17.21 & - & 60cm & FLI\\
2013 Aug 09 & 56514.427 & 14.05 & 15.21 & 16.10 & 17.30 & - & 60cm & FLI\\
2013 Sep 04 & 56540.376 & 14.03 & 15.33 & 16.25 & 17.41 & - & Sch & FLI\\
2013 Sep 06 & 56542.440 & 14.01 & 15.24 & 16.14 & 17.29 & - & Sch & FLI\\
2013 Sep 08 & 56544.406 & 14.06 & 15.02 & 16.04 & 17.16 & - & 2m & VA\\
2013 Sep 11 & 56547.471 & 14.01 & 15.20 & 16.13 & 17.33 & - & 60cm & FLI\\
2013 Sep 14 & 56550.302 & 14.08 & 15.33 & 16.30 & 17.30 & - & 60cm & FLI\\
2013 Sep 17 &	56553.364	& 14.00 &	15.26 &	16.20 &	17.35 &	- &	1.3m &	ANDOR\\
2013 Oct 11 & 56577.445 & 13.98 & 15.25 & 16.17 & 17.08 & - & 60cm & FLI\\
2013 Oct 12 & 56578.461 & 14.02 & 15.34 & 16.18 & 17.40 & - & 60cm & FLI\\
2013 Nov 07 & 56604.423 & 13.91 & 15.12 & 15.99 & 17.15 & - & 60cm & FLI\\
2013 Dec 29 & 56656.307 & 14.00 & 15.25 & 16.17 & 17.28 & - & Sch & FLI\\
2014 Jan 23 & 56681.282 & 13.95 & 15.08 & 16.02 & 17.17 & - & Sch & FLI\\
2014 Feb 05 & 56694.593 & 14.05 & 15.08 & 16.08 & 17.26 & - & 2m & VA\\
\end{longtable}

\end{document}